# Is Dark Energy Falsifiable?

Carl H. Gibson
Departments of MAE and SIO
Center for Astrophysics and Space Sciences
University of Cal. San Diego
La Jolla, CA 92093-0411
cgibson@ucsd.edu

Rudolph E. Schild
Harvard-Smithsonian Center for Astrophysics
60 Garden Street, Cambridge, MA 02138
rschild@cfa.harvard.edu

## Abstract

Is the accelerating expansion of the Universe true, inferred through observations of distant supernovae, and is the implied existence of an enormous amount of anti-gravitational dark energy material driving the accelerating expansion of the universe also true? To be physically useful these propositions must be falsifiable; that is, subject to observational tests that could render them false, and both fail when viscous, diffusive, astro-biological and turbulence effects are included in the interpretation of observations. A more plausible explanation of negative stresses producing the big bang is turbulence at Planck temperatures. Inflation results from gluon viscous stresses at the strong force transition. Anti-gravitational (dark energy) turbulence stresses are powerful but only temporary. No permanent dark energy is needed. At the plasma-gas transition, viscous stresses cause fragmentation of plasma proto-galaxies into dark matter clumps of primordial gas planets, each of which falsifies dark-energy cold-dark-matter cosmologies. Clumps of these planets form all stars, and explain the alleged accelerating expansion of the universe as a systematic dimming error of Supernovae Ia by light scattered in the hot turbulent atmospheres of evaporated planets surrounding central white dwarf stars.

## 1. Introduction

If true, observations that the rate of expansion of the universe ceased to decelerate and began to accelerate about a billion years ago is important enough to justify the 2011 Nobel Prize in Physics for the three recipients Perlmutter, Reiss and Schmidt. However, the inference of accelerated expansion of the universe from Supernovae Ia dimming is immediately falsified if the dark matter of galaxies is not cold dark matter halos but primordial gas planets in clumps (Gibson, 1996; Schild, 1996). The accelerated rate of expansion proposition is based on the faulty assumption that the observed dimming of Supernovae Ia events at redshift z about 0.46 reflects accelerating distance to the events. But 20% systematic dimming error due to light scattering from large evaporated planet atmospheres at Oort cloud distances from overfed central carbon stars is exactly what is expected from hydrogravitational dynamics (HGD cosmology). When the star converts too many planets to carbon it explodes and part of the light may or may not be dimmed by the ambient planets atmospheres. The second proposition that a huge amount of permanent anti-gravitational (dark energy) material ($\Lambda$ = 0.7) exists to drive the accelerating expansion is even more unwarranted and questionable.

Both propositions are false because the standard cold dark matter hierarchical clustering model $\Lambda$CDMHC is false, as proposed by Gibson (1996) and confirmed by the quasar microlensing observations of Schild (1996). The reasons are quite easy to understand and test. Because $\Lambda$CDMHC fails to include modern fluid mechanics (and even not-so-modern fluid mechanics concepts like kinematic viscosity and diffusivity) it fails to recognize that the dark matter of galaxies is clumps of primordial-gas planets, 30,000,000 planets per star. The planets have Earth mass and larger, and appear in trillion planet clumps with the mass and density of a globular star cluster. All stars form in these clumps as the planets merge to form larger planets and eventually stars. Overfeeding of a star by planets leads to white dwarf formation, evaporation of ambient planets to form planetary nebulae, and Supernovae Ia events. Partially evaporated planets that surround white dwarf stars of planetary nebulae have complex refractive index fluctuations in their atmospheres that may dim SnIa events if they are on the line of sight to the event (Schild & Dekker 2006).

The fluid mechanical assumptions of $\Lambda$CDMHC cosmology generally reflect those of James Jeans 1902. Jeans' 1902 assumptions are mathematically convenient, but are now quite obsolete and misleading, and should be replaced by hydrogravitational dynamics HGD cosmology, as illustrated by Figure 1. We suggest the observational discovery of the 20[th] century was that of Schild (1996), not the Supernovae Ia dimming observations. From the twinkling frequency of quasar images, Schild was able to infer that the missing mass of the lensing galaxy is dominated by Earth-mass planetary objects, not stars. The observations have been repeatedly confirmed. From the brightness of the mirage images of the quasar, it follows that the planets are in clumps, as predicted by HGD cosmology.

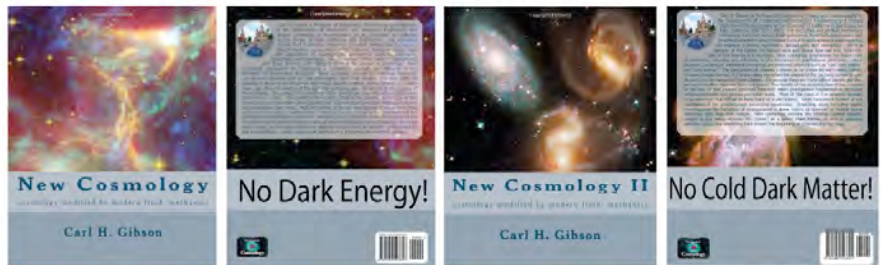

Figure 1. Warnings about false predictions of $\Lambda$CDMHC "concordance" cosmology (CC) have been made repeatedly since 1996 when hydrogravitational dyanamics HGD cosmology was introduced by Gibson (1996) and Schild (1996). Cosmological flows generally are not collisionless, and are not frictionless and ideal as assumed.

Several collaborations claim their microlensing observations falsify planetary mass objects as the missing mass of galaxies (MACHO, OGLE, EROS), but all underestimate the total mass from the incorrect assumption that the planets are uniformly distributed in the galaxy rather than concentrated in protoglobularstarcluster (PGC) clumps.

Attempts to explain the observations of supernovae dimming by standard theories of physics and the cosmological constant have failed spectacularly. Particle physics and quantum gravity analysis without fluid mechanics give cosmological constant $\Lambda$ values too large by a factor of between one hundred and sixty orders of magnitude (Witten 2008). This result by itself would seem to be an adequate falsification of dark energy, which raises the eponymous question of whether any theory or any observations can accomplish this task among faithful members of the concordance cosmology community.

In the following, we review the theory, the observations, and present a summary and conclusions.

## 2. Theory

The standard concordance cosmology (CC) theory of gravitational structure formation in the universe assumes collisionless potential flow for mathematical and numerical convenience. This leaves the mechanism of the big bang a mystery, and requires the invention of cold dark matter (CDM): an unknown collisionless material that condenses gravitationally and hierarchically clusters (HC) to form CDM halos that increase with size and mass over time. Observations of galaxies at great distances require early structure formation, starting during the plasma epoch between $10^{11}$ seconds, when mass exceeds energy, and $10^{12}$ seconds, when superclusters fragment and the baryonic density is $\rho_0 = 5 \times 10^{-17}$ kg m$^{-3}$. Plasma turns to gas at $10^{13}$ s (0.3 Myr). The Jeans length scale $L_J = V_S/\tau_g$ for the plasma is larger than the horizon scale $L_H = ct$ during this period, so structure formation is impossible by the Jeans 1902 criterion, where $V_S$ is the speed of sound $c/3^{1/2}$, c is the speed of light, t is the time since the big bang, $\tau_g$ is the gravitational free fall time $(\rho G)^{-1/2}$, $\rho$ is density and G is Newton's gravitational constant. By this scenario, CDM halos grow by HC merging of CDM seeds formed in the plasma epoch to form gravitational potential wells that gravitationally collect the baryonic plasma. A period of 400 Myr, known as the dark ages, is required to form the first star. The last structures to form by CDMHC (CC) cosmology are the largest; that is, the superclusters and superclustervoids. From CC they both should be the same size, about $10^{23}$ meters.

The easiest way to falsify the dark energy and cold dark matter concepts (Gibson 2011) is to recognize that life is quite impossible without miracles in CC cosmologies but inevitable and early in HGD cosmology (Gibson, Wickramasinghe and Schild 2011). Life fails in CC cosmology because, after 400 million years of CDMHC dark ages, the first planets to appear are too rare, too cold, and too isolated, and have no liquid water oceans. Life succeeds in HGD cosmology because its planets merge to make all stars, the stars make chemicals the planets convert to water and life. The planets carry the seeds of life with them as they merge to form larger planets and stars, explaining the cometary panspermia mechanism pioneered by Fred Hoyle and Chandra Wickramasinghe.

Hydrogravitational dyamics (HGD) cosmology predictions are very different from those of CDM cosmology, particularly for the largest scale objects, as shown in Figure 2. Granett, Neyrinck and Szapudi (2011) claim the linear ISW effect is a direct signal of dark energy. Figure 2 is modified from a figure on a website created by these authors. According to HGD cosmology however, the largest scale objects like protosuperclusters and protogalaxies are all formed by condensations and fragmentations within the plasma epoch time period shown as a dashed circle on the left of Fig. 2. Protosuperclustervoid growth occurs as a rarefaction wave at the speed of sound $c/3^{1/2}$ to establish the sonic peak of the CMB spectrum at the length scale $tc/3^{1/2} \sim 10^{21}$ m, as observed. The superclustervoids continue to grow by the expansion of space and the forces of gravity to their present observed size near $10^{25}$ m, according to HGD cosmology. This is much larger than the maximum size $10^{23}$ meters expected from CDMHC cosmologies.

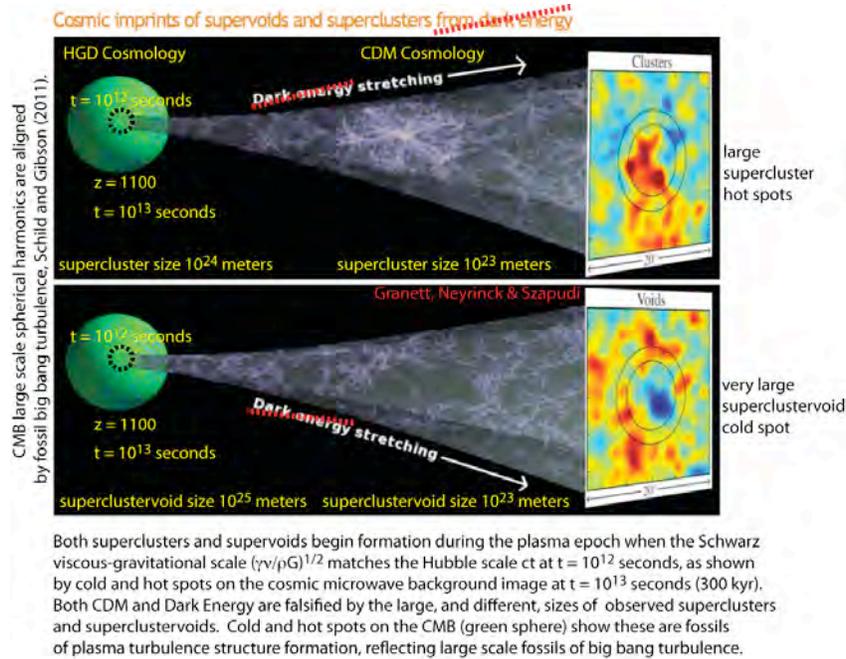

Figure 2. Predictions of HGD (HydroGravitational Dynamics) and ΛCDMHC "concordance" (CC) cosmologies differ strongly at supercluster and superclustervoid scales, Schild and Gibson (2011).

As shown in Fig. 2, observations of large scale structures strongly depart from the expectations of CC cosmologies. Observed superclusters and superclustervoids are too large and the superclustervoids are significantly larger than the superclusters, as expected from HGD cosmology. Growth of large structures occurs much too late in CC cosmologies to match the observations. Dark energy and cold dark matter concepts are falsified.

Figure 3 shows that a $10^{25}$ meter (300 Mpc) void has been measured by radio telescopes (Rudnick et al. 2008). The position on the sky of the measured supervoid is that expected from the CMB great cold spot on the cosmic microwave background temperature anisotropy map determined by the WMAP team (Cruz et al. 2006).

Efforts to detect effects of dark energy in supervoid regions using the integrated Sachs Wolfe ISW effect are reported by Nadathur et al. (2011). The result reported in the abstract falsifies both dark energy and cold dark matter: "If the observed signal is indeed due to the ISW effect then huge, extremely underdense voids are far more common in the observed universe than predicted by ΛCDM".

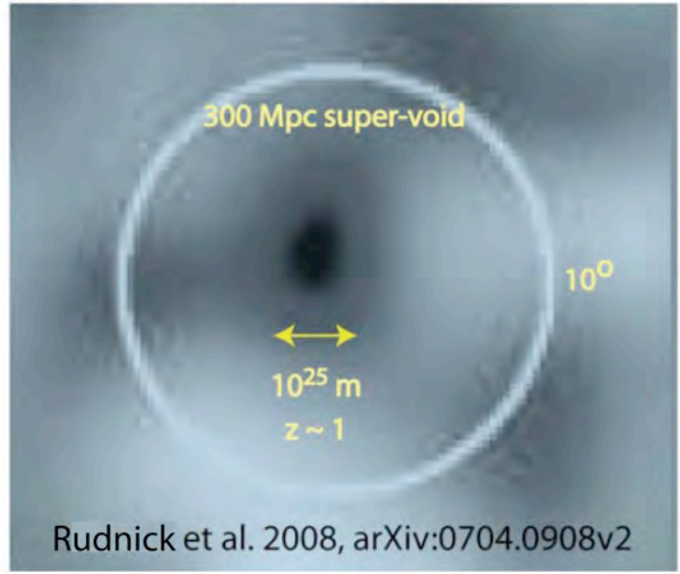

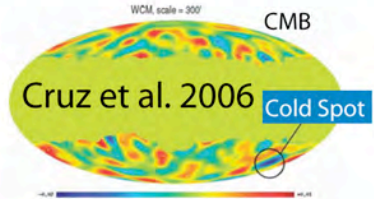

Figure 3. Rudnick et al. 2008 show radio telescope evidence of a 300 Mpc ($10^{25}$ meter) diameter supervoid associated with the great cold spot observed on the cosmic microwave background, as illustrated by Cruz et al. 2006. Such a large empty region falsifies dark energy and cold dark matter hierarchical clustering scenarios. An average growth rate about 10% of the speed of light is required, as expected from HGD cosmology.

To create the magnitude and angular size of the WMAP cold spot requires a ~ 140Mpc radius completely empty void at z < 1 along this line of sight. This is far outside the current expectations of concordance cosmology, and adds to the many "anomalies" seen in the CMB such as the preferred "axis of evil" direction (Schild and Gibson 2011).

Figure 4 compares time lines of CC cosmology (from STSci-PRC-01-09) and HGD cosmology. As shown, the two cosmologies are very different, almost orthogonal, so it is easy to use observations to falsify one or the other. The easiest case to make is from astro-biology, where evidence now clearly shows an abundance of extraterrestrial life throughout the Universe has a common organic chemical basis (Gibson, Wickramasinghe and Schild 2011) that is easily explained by the large number of communicating primordial gas planets of HGD cosmology as the dark matter of galaxies.

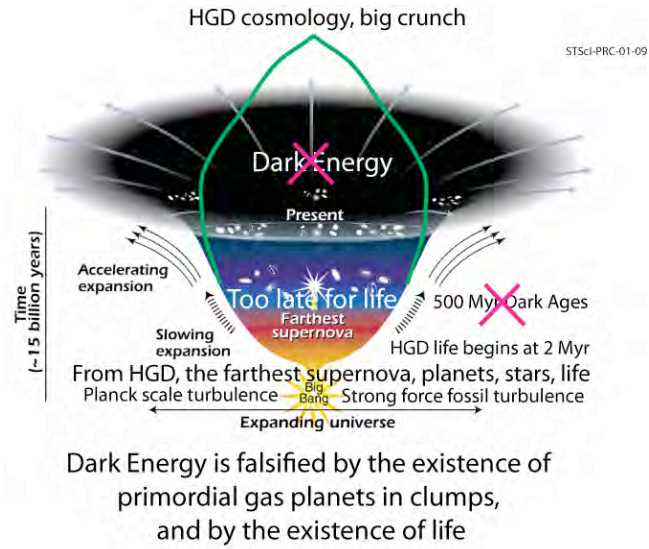

Figure 4. Comparison of theoretical predictions of HGD cosmology versus the present standard ΛCDMHC (concordance) cosmology including Accelerating Expansion driven by anti-gravity Dark Energy. Because the two cosmologies are so different, observations should be able to falsify either one or the other in many ways. Life is quite impossible by concordance cosmology because of the long ~500 Myr dark ages period needed before stars and planets can appear, so life falsifies CC cosmology. Huge supervoids and primordial gas planets in clumps predicted by HGD cosmology as the galaxy dark matter also falsify CC cosmology.

As shown in Fig. 4, both HGD and CC (Dark Energy) cosmologies begin with a big bang, 13.7 Gyr before the present time. Turbulent anti-gravity forces of HGD cosmology are needed to overcome $10^{51}$ g gravitational accelerations that arise at Planck scales ($10^{-35}$ meters) during the big bang (Gibson 2010). However, these turbulent vortex stretching forces vanish when the turbulent fireball is damped by gluon viscous forces at $10^{-27}$ seconds when inflation begins. Entropy produced by turbulent friction results in the prediction of a closed universe by HGD, with a big crunch in the future rather than the accelerated expansion rate driven by dark energy forces.

Convincing arguments and observations support the existence of a cosmological big bang origin of the universe 13.7 billion years before the present time. How did it begin? Only turbulence vortex dynamics can explain the enormous stresses required at Planck scales to overcome gravity, and the resulting anisotropies observed in the largest scale spherical harmonics of the cosmic microwave background, Schild and Gibson (2011).

A fatal flaw in CC-cosmology is the consistent neglect of kinematic viscosity $\nu$ in determining the formation of gravitational structures, Gibson (1996). Gluon viscosity produced inflation from negative stresses, and $10^{97}$ kg of mass-energy at $10^{-27}$ seconds. Photon viscosity produced the first structures of the plasma epoch with the density of globular star clusters at $10^{12}$ seconds, with galaxies as the smallest fragments at $10^{13}$ seconds. Primordial gas viscosity predicts Earth mass planets at this time of plasma to gas transition. A second (Jeans) scale of fragmentation of the gas results from the large difference between the sound speed and light speed, causing fragmentation of globular star cluster GC mass clumps of primordial planets with GCs in all galaxies identical size $4 \times 10^{17}$ m, identical mass ~$10^6$ solar and identical density $\rho_0$.

HGD cosmology consists of a sequence of phase transitions starting with a turbulent big bang permitted by Planck temperatures of a big crunch, where only Planck particles and antiparticles exist and the only length scale is $10^{-35}$ meters. Inertial vortex forces of the turbulence easily provide sufficient antigravitational stresses to overcome the Fortov-Planck "dark energy" $\Lambda_{FP} = c^7 h^{-1} G^{-2} = 10^{113}$ Pa, where c is the speed of light, h is Planck's constant and G is Newton's constant. Gluon viscous stresses explain inflation. Photon viscous stresses explain protosuperclustervoids and protogalaxies fragmented during the plasma epoch. Protogalaxies are more massive than protoplanets because the plasma photon viscosity is $10^{13}$ larger than the gas viscosity at transition.

Contrary to CC-cosmology, the enormous diffusivity $D_{NB}$ of the non-baryonic component of the universe compared to the viscosity and diffusivity of the baryonic components causes their diffusive separation at galaxy scales. Cold dark matter, if it existed, could not condense or clump or hierarchically cluster because $D_{NB} \gg D_B$. Instead this most massive component of the universe is nearly irrelevant to gravitational structure formation, and diffuses to form large-scale halos of galaxy clusters and superclusters.

Evidence of extraterrestrial biology is particular hostile to the validity of CC-cosmology, which relies on the condensation of cold dark matter in clumps during the plasma epoch that must cluster to form gravitational potential wells that collect the baryons and eventually form the first stars and a few planets after half a billion years at temperatures much too cold to foster life.

## 3. Observations

The most important observations relevant to cosmology were not those of supernova dimming leading to the 2011 Nobel Prize in Physics and dark energy but of quasar microlensing by a galaxy, giving the unexpected result that the dominant point mass objects of the galaxy were not stars but planets (Schild 1996). This enormously massive population of planets was correctly interpreted by Schild as the missing mass of the galaxy. The discovery revolutionizes cosmology and all of astronomy. Stars must come from the planets, rather than the planets coming from stars. Since all stars must be formed in dense clumps of planets, it seems obvious that the planets surrounding such stars could cause systematic dimming errors in supernova brightness that could be misinterpreted as evidence of dark energy driven accelerated expansion of the universe. This source of error has not been taken into account by CC-cosmology.

From the outset, it was recognized that dark matter dimming of the observed supernovae provided a satisfactory explanation for the observed supernovae without appeal to a hypothesized dark energy. When Goobar, Bergstrom and Mortsell (2002) showed that a simple "self-replenishing dust" model easily matched the supernova data, the absorbing matter was considered to be dust grains with larger sizes that would not give a reddening signal long associated with interstellar Dust in the Galaxy. In turn, Schild and Dekker (2006) examined the refraction effects that the planetary mass repositories of the Baryonic dark matter would have. Such objects acting as spherical lenses would refract the light of a background supernova, and diminish the supernova brightness without leaving any color signature. The refracted light would be lensed into a different direction of space, and contribute to the night sky background. These effects were not considered at the time that Dark Energy was being discussed (Riess, 2004), even though the planetary mass population had been recognized in quasar microlensing observations years previously (Schild 1996).

Nor would it be surprising that the population of "grey dust" absorbers hypothesized by Goobar et al. (2002) would have the evolutionary properties expected. Grey dust is presented as solid particles from which the smallest, reddening, particles have been systematically removed by unknown processes. HGD cosmology explains both the lack of small grey dust particles and the intermittency of SnIa dimming, as observed.

The proposed absorbers fitted to supernova brightness as demonstrated by Goobar et al. (2002) and illustrated in their Fig. 3 (bottom) are required to have an extinction law parameterized as $A = A_0(1+z)^3$. We prefer to think of this law as an extinction decreasing with advancing time since $z = 0.5$, and constant at earlier times. Such an extinction model in a universe with $\Lambda_M = 1$ then gives a satisfactory fit to the supernova brightness data (ie, without needing any dark energy) up to the $z = 1.7$ upper limit of measured supernova brightness and redshift.

In other words, a simple $\Lambda_M$ matter dominated universe ($\Lambda = 0$) with the modeled simple grey dust extinction law fits acceptably to the supernova observations.

Recall that the planetary mass population of condensed objects (MACHOs) would have formed at plasma-to-gas transition, 300,000 years after the big bang in clumps with the baryonic density existing at the time of first fragmentation 30,000 years. Within a few million years they would have condensed to a clumped, micro-brown-dwarf, planetary population with hot central cores of star formation within the (protoglobularstarcluster PGC) clumps radiating away the heat of gravitational binding. In this process of cooling throughout the 13.7 Gyr history of the universe, they would lag behind the universe's background temperature, measured in the local universe to be 2.7 degrees Kelvin. At earlier times, when the falling background temperature passed through the Hydrogen triple point temperature of 13.8 degrees Kelvin, hydrogen ice would have begun forming, and fallen toward the planetary surface, but initially re-evaporated, creating an active weather system. The central temperatures and radii would depend on the mass and chemical composition of the planetary condensation. Because of the large central reservoir of planetary latent heat, the radiative cooling would significantly lag the universe's background radiation cooling.

Thus the primordial planet atmospheres would have been continually shrinking from their initial $\sim 10^{14}$ m hot primordial gas size, and contracting dramatically to $\sim 10^7$ m after the outer layers had cooled to freezing below the hydrogen triple point at $\sim 24$ Myr.

A second requirement of the Goobar extinction model is that the extinction should be approximately achromatic, as specified by a brightness to B-V color ratio, R, of 5.5 or higher. As noted previously, a process of refraction by the primordial planet hydrogen atmospheres acting as spherical lenses would achromatically refract light out of the supernova image and achromatically cause the observed dimming properties with the required time evolution property. We avoid the term "grey extinction" because the process of refraction would deflect the supernova light to a different direction in the universe and contribute to the cosmic background light. These topics have been discussed in greater detail in Schild (2004) and Schild and Dekker (2006).

While this is reasonably understood qualitatively, no detailed modeling following the thermal and physical history of a rogue planet has yet been constructed. While some brown dwarf modeling has been undertaken by Baraffe, Chabrier, Barman (2008) to compare to existing observations of brown dwarf objects, no comprehensive evolutionary track is yet established.

An important development, however, is available from the identification of these primordial planets (micro-brown dwarfs, $\mu$BD) as discussed by Nieuwenhuizen, Schild and Gibson (2011) and Schild and Dekker (2006), with the long-observed Lyman-alpha clouds. The existence and nature of these clouds has never been understood (Rausch 1991) because they are recognized in quasar spectra to large redshifts, and therefore must have existed at early epochs of the Universe, but it is difficult to imagine from the standard model how they could survive without simply diffusing away. However if the hydrogen clouds producing the Lyman alpha forest absorption lines are internally condensed as primordial planet MACHOs, then long lifetimes are understandable. PGC (protoglobularstarcluster) clouds of planets persist indefinitely in metastable equilibrium, frozen until agitated within their clumps. In this picture, we propose that the Earth was manufactured by a complex planet merger process over the 13.7 billion year history of the universe. It first condensed gravitationally just after plasma neutralization 300,000 years after the big bang, and then collected up its endowment of heavy elements from the first stars from merging hot gas planets and their cores after the first supernova events. At this phase, interstellar dust grains such as iron and silicon oxides meteored into the enormous hydrogen atmospheres, reducing the metals to make a growing iron-silicon core surmounted by oceans of water (Gibson, Wickramasinghe and Schild 2011). Mergers of such planets to form larger planets and finally stars constantly recycled the materials, and shared chemical and biological information as it accumulated and evolved in the planet water oceans.

If the Lyman alpha clouds can be identified with frozen primordial planets intermittently surrounded by their hydrogen envelopes depending on friction and ambient radiation, then the statistics of their number are relevant to absorption of background supernovae light. As discussed above, the primordial planets would be cooling and converting their hydrogen envelopes to hydrogen ices that precipitate to their surfaces, thus shrinking the objects and creating an evolving, apparently diminishing population. The observed density of Lyman-alpha clouds has been long recognized to follow a power law distribution with exponent approximately 2.8. In other words, $N(z) \sim (1+z)^{2.8}$. And it was recognized that the "self replenishing dust" model with absorption scaling as $(1+z)^3$ up to $z = 0.5$ and constant to higher redshifts adequately described the Riess et al (2004) supernova brightness-redshift law, without invoking dark energy (Goobar et al. 2002).

We conclude that knowledge already discussed can adequately explain the observed properties of the SN Ia brightness-distance law that was interpreted a decade ago as proving the existence of a repulsive Dark Energy. Although other lines of evidence have been cited as demonstrating such a Dark Energy, they are always superimposed upon the ΛCDM model of Dark Matter, which is presently failing as its predictions are being compared to observations. Particularly troubling are ΛCDM predictions of the upper limits to scales of cosmic voids, and predictions of unobserved subhalos in the Galactic Halo.

Clearly so many planets, $\sim 10^{80}$ from the big bang, must be composed of primordial gas, and must be relevant to the formation of life. Spectroscopic examination of the intergalactic dust reveals it is almost certainly the result of life on extraterrestrial dark matter planets. Observations support the Gibson, Wickramasinghe and Schild (2011) claim that life began soon after the plasma to gas transition, at 2 Myr rather than after 400 Myr of dark ages, as shown in Fig. 4. Figure 5 shows recent evidence of extraterrestrial life from infrared spectra from the interstellar medium (Kwok and Zhang 2011).

The existence of planets in clumps as the dark matter of galaxies certainly calls the dark energy hypothesis into question. Biological evidence that life began at primordial times on primordial planets cannot possibly be reconciled with CC-cosmology, the accelerated expansion of the universe, and the dark energy explanation of the accelerated expansion.

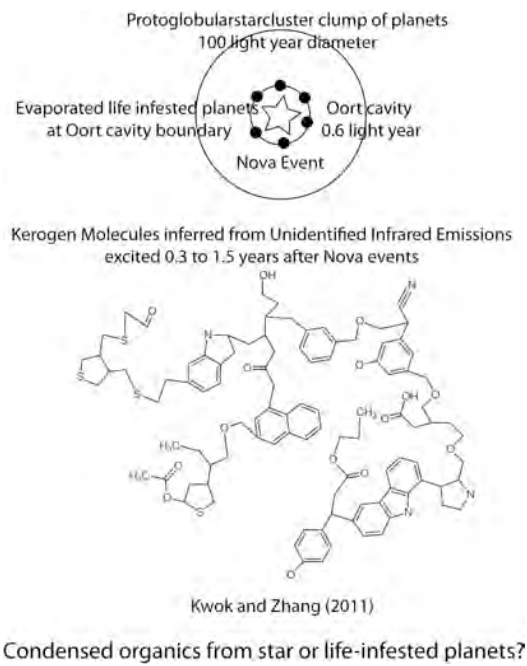

Figure 5. Complex chemicals inferred from infrared emissions following Nova events begin at hundred day times that reflect the size of Oort cavities produced when stars are formed from planets in PGC clumps.

Recent infrared telescopic evidence of complex organic molecules similar to kerogen is provided by Kwok and Zhang (2011). Kerogen is the part of coal shale that cannot be dissolved, versus bitumen, which can. Bright infrared emissions are observed starting about 0.3 years after star Nova events, corresponding to the 0.1 pc ($3 \times 10^{15}$ m) size of the Oort cavity produced when a star is formed in a protoglobularstarcluster PGC clump of dark matter planets, shown at the top of Fig. 5, and by dashed circles in Fig. 6.

The UIE (unidentified infrared events) have been identified as "condensations" of organic chemicals somehow emitted by the Nova stars. It is physically impossible for such complex organic chemicals to be produced, or survive, a Nova event, but it is easy to explain these observations from HGD cosmology. The black dots at the top of Fig. 5 represent the same evaporated planets that falsify dark energy in CC-cosmology by producing a systematic dimming error in supernovae Ia events. Identification of kerogen molecules in Nova spectra, bottom Fig. 5, further falsifies the dark energy hypothesis.

Strong evidence in support of primordial gas planets in clumps as the source of all stars by planet mergers is now available from the Herschel space observatory study of infrared dense cores IRDC in the galactic plane (Wilcock et al. 2011), Figure 6. The sample image is from a protoglobularstarcluster PGC clump (dense molecular cloud) detected at 8 microns by Spitzer space telescope as a dark region. Fig. 6 from Herschel is from emission at 150 μ meter wavelength, with temperatures near 14 K, the triple point of hydrogen.

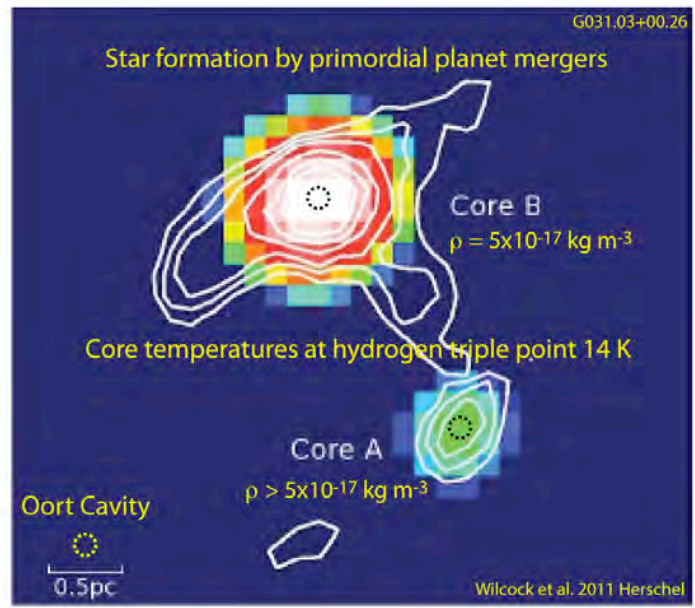

Figure 6. Star formation by mergers of frozen primordial gas planets detected by the Herschel space observatory (Wilcock et al. 2011, Fig. 2). The measured density of Core B is precisely the fossilized value $\rho_0 = 5 \times 10^{-17}$ kg m$^{-3}$ expected for proto-globular-star-cluster PGC clumps of dark matter planets, the baryonic density at $10^{12}$ seconds, from HGD cosmology. The measured density at Core A becomes larger than $\rho_0$ as planets merge toward stellar mass. Radiation pressure outward limits the condensation of planets at the 0.1 pc scale of an Oort cavity (dashed circles). The average density in a galaxy is $\sim 10^{-4} \rho_0$. The separation of dark matter planets is $3 \times 10^{-3}$ pc ($10^{14}$ m).

As the planets merge to form a star they evaporate at the triple point of frozen hydrogen

13.8 K, which is the temperature measured at the two dense cores A and B. Total masses of the cores are about 100 solar, as expected from the primordial density of PGC clumps following first plasma fragmentation at 30 Kyr according to HGD cosmology. What is being formed at Cores A and B in Fig. 6 are protostellar discs and Oort cavities around binary stars. These eventually explode as a SnIa from continued accretion of planets.

## 4. Summary and Conclusions

It is very inconvenient for the concepts of dark energy and cold dark matter to be falsified. This requires a revolution in the basic concepts of how gravitational structures are formed at every stage of cosmology, starting with the big bang. Expensive numerical simulations assuming potential flows of collisionless fluids become suspect. Star formation theory must be completely revised if the dark matter of galaxies is Earth mass planets in protoglobularstarcluster clumps, as observed (Schild 1996) and predicted (Gibson 1996), since stars must be formed by mergers of these planets. Planet formation theories must also be revised, since stars are formed by planets, not planets formed by stars. The remarkable stability of trillion-planet-clumps of primordial planets must be explored. Enormous empty void regions are observed that cannot be explained by the standard concordance cosmology but is natural in HGD cosmology.

Evidence of extraterrestrial life also falsifies the hypothesis that the universe is expanding at an accelerating rate due to antigravity effects of a permanent dark energy Λ. It is impossible to deny that life exists on Earth. It is becoming impossible to deny that life would not exist on Earth without Hoyle/Wickramasinghe cometary panspermia. Combining cometary panspermia with HGD cosmology shows that life began in a biological big bang almost immediately after the cosmological big bang, and was spread homogeneously throughout the dark matter planet-clumps produced, taking with them genetic evidence of the same, and primordial, last universal common ancestor (LUCA) in every organism (Gibson, Wickramasinghe and Schild 2011). New telescopes, and now new microscopes, can easily and precisely discriminate between predictions of the two cosmological models, which are very different. At some point the rising flood of evidence supporting HGD-cosmology and contradicting CC-cosmology and dark energy becomes overwhelming. We suggest that point has long since been passed.

### 6. References

Baraffe, I. Chabrier, G. & Barman, T. Structure and Evolution of Super-Earth to Super-Jupiter Exoplanets. 1. Heavy Element Enrichment in the Interior, A&A, 482, 315.

Cruz, M., Tucci, M., Martinez-Gonzalez, E., & Vielva, P. The non-Gaussian cold spot in Wilkinson Microwave Anisotropy Probe: significance, morphology and foreground contribution, MNRAS, 369, 57, 2006.

Gibson, C.H., Turbulence in the ocean, atmosphere, galaxy and universe. Appl. Mech. Rev. 49, no. 5, 299–315, 1996.

Gibson, C. H., Turbulence and turbulent mixing in natural fluids, Physica Scripta, Turbulent Mixing and Beyond 2009, T142 (2010) 014030 doi: 10.1088 /0031-8949 /2010 /T142/0140302010, arXiv:1005.2772v4, 2010.

Gibson, C. H., Does Cometary Panspermia Falsify Dark Energy?, Journal of Cosmology,

16, 7000-7003, 2011.

Gibson, C. H. and R. E. Schild, Clumps of hydrogenous planetoids as the dark matter of galaxies, arXiv:astro-ph/9908335v2. Submitted to ApJ in 1996a.

Gibson, C. H. and R. E. Schild, Quasar-microlensing versus star-microlensing evidence of small-planetary-mass objects as the dominant inner-halo galactic dark matter, arXiv:astro-ph/9908335v2. Submitted to ApJ in 1996b.

Gibson, C. H., T. M. Nieuwenhuizen and R. E. Schild, Why are so many primitive stars observed in the Galaxy halo?, Journal of Cosmology, 16, 6824-6831, 2011.

Gibson, C. H., N. C. Wickramasinghe, and R. E. Schild, The Biological Big Bang: The First Oceans of Primordial Planets at 2-8 Myr Explains Hoyle – Wickramasinghe Cometary Panspermia and a Primordial LUCA. Journal of Cosmology, 16, 6500-6518, 2011.

Goobar, A. Bergström, L. & Mörtsell, E. Measuring the properties of extragalactic dust and implications for the Hubble diagram, Astron. & Astrophys. 384, 1, 2002.

Granett, B. R., Neyrinck, M. C. and Szapudi, I., An Imprint of Super-Structures on the Microwave Background due to the Integrated Sachs-Wolfe Effect, The Astrophysical Journal, 683: L99–L102, 2008.

Joseph, R. and N. C. Wickramasinghe, Genetics Indicates Extra-Terrestrial Origins for Life: The First Gene, Did Life Begin in the "Moments" Following the Big Bang? Journal of Cosmology, 16, 6832-6861, 2011.

Kwok, Sun and Yong Zhang, Mixed aromatic–aliphatic organic nanoparticles as carriers of unidentified infrared emission features, Nature, Oct. 26, doi:10.1038/nature10542, 2011.

Nadathur, S., S. Hotchkiss, and S. Sarkar, The integrated Sachs-Wolfe imprints of cosmic superstructures: a problem for ΛCDM, arXiv:1109.4126v1 [astro-ph.CO] 19 Sep 2011.

Nieuwenhuizen, Theo M.; Schild, Rudolph E.; Gibson, Carl H. Do micro brown dwarf detections explain the galactic dark matter? ArXiv:1011.2530, Journal of Cosmology, 15, 6017-6029, 2011.

Rauch, M., The Lyman Alpha Forest in the Spectra of QSOs, Ann Rev, Astron. & Astrophys, 36, 267, 1998.

Rudnick, L., S. Brown and L. R. Williams, Extragalactic Radio Sources and the WMAP Cold Spot, The Astrophysical Journal, 671, 40-44, 2007.

Schild, R. E., Microlensing variability of the gravitationally lensed quasar Q0957+561 A,B. ApJ 464, 125, 1996.

Schild, R. E., Some Consequences of the Baryonic Dark Matter Population, Article for the 2004 Edinburgh Dark Matter Conference, (2005)

Schild, R. E. & Dekker, M. The Transparency of the Universe Limited by Lyman-alpha Clouds, Astron. Nachr. 327, 729, (2006)

Schild, R. E and C. H. Gibson, Goodness in the Axis of Evil, Journal of Cosmology, 16, 6892-6903, 2011

Wilcock, L. A. et al., The initial conditions of high-mass star formation: radiative transfer models of IRDCs seen in the Herschel Hi-GAL survey, arXiv:1101.3154v1 [astro-ph.GA], accepted A&A, 2011.

Witten, Edward, stsci 2008 spring symposium .http: //www .stsci .edu /institute /itsd /information /streaming /archive /SpringSymposium2008.